\documentclass[a4paper,11pt]{article}
\pdfoutput=1 

\usepackage{jcappub} 
\usepackage{indentfirst}
\usepackage{appendix}
\usepackage{multirow}
 
\def\blue#1{{\textcolor{blue}{#1}}}
\usepackage{array}
\usepackage{enumerate}
\usepackage{graphicx}
\usepackage{dcolumn}
\usepackage{CJKutf8}
\usepackage{bm}
\usepackage{amssymb}
\usepackage{latexsym}
\usepackage{booktabs}
\usepackage{amsmath}
\usepackage{subcaption}
\usepackage{url}

\begin{document}
\title{Prospects for joint multiband detection of intermediate-mass black holes by LGWA and the Einstein Telescope}

\author[a]{Yue-Yan Dong,}
\author[a]{Ji-Yu Song,}
\author[a]{Jing-Fei Zhang}
\author[a,b,c,1]{and Xin Zhang\note{Corresponding author.}}
\affiliation[a]{Key Laboratory of Cosmology and Astrophysics (Liaoning Province), College of Sciences, Northeastern University, Shenyang 110819, China}
\affiliation[b]{Key Laboratory of Data Analytics and Optimization for Smart Industry (Ministry of Education), Northeastern University, Shenyang 110819, China}
\affiliation[c]{National Frontiers Science Center for Industrial Intelligence and Systems Optimization, Northeastern University, Shenyang 110819, China}

\emailAdd{dongyueyan@stumail.neu.edu.cn, songjiyu@stumail.neu.edu.cn, jfzhang@mail.neu.edu.cn, zhangxin@mail.neu.edu.cn}

\abstract{Gravitational-wave (GW) detection offers a novel approach to exploring intermediate-mass black holes (IMBHs). The GW signals from IMBH mergers mainly fall in the decihertz frequency band. The lunar-based GW detector, the Lunar Gravitational-Wave Antenna (LGWA), exhibits high sensitivity in this band, making it particularly well-suited for detecting IMBHs. However, for lower-mass IMBHs, the late inspiral and merger signals enter the sensitive frequency range of ground-based GW detectors.
In this work, we aim to explore how multi-band observations with LGWA and the third-generation ground-based GW detector, the Einstein Telescope (ET), can contribute to detecting the population of IMBHs. 
We consider three population distribution cases of IMBHs, including two population models based on astrophysical motivations and a uniform distribution, and compute the signal-to-noise ratios for LGWA, ET, and their combination to directly compare their capabilities in detecting IMBH mergers.
Our results suggest that LGWA possesses strong detection capability for high-mass IMBH mergers. At redshift $z = 1$, LGWA's detection rate for IMBH binaries with primary masses above $5 \times 10^4~M_\odot$ is largely insensitive to orbital inclination and mass ratio. In contrast, ET is more suited for detecting IMBH binaries with primary masses below $10^3~M_\odot$. The multi-band observation of LGWA and ET possesses strong detection capabilities across the full IMBH mass spectrum. Furthermore, we find that the multi-band detection can significantly and effectively recover the IMBH population distributions.
In summary, we conclude that the multi-band observations of LGWA and ET will provide powerful detection capabilities for IMBHs and are expected to significantly enhance our understanding of this important yet still poorly observed class of black holes.}

\maketitle

\section{Introduction}\label{sec:intro}

Intermediate-mass black holes (IMBHs), with masses between $10^2$ and $10^5\ M_\odot$, bridge the gap between stellar-mass black holes and supermassive black holes, and could play a critical role in the formation and evolution of galaxies, as they could be the seeds of supermassive black holes \cite{Greene:2019vlv, Madau:2001sc, Silk:2017yai, Natarajan:2020avl}. Moreover, IMBHs are expected to have profound implications for both astrophysics and cosmology, serving as potential sources of tidal disruption events \cite{Rosswog:2008ie, MacLeod:2015jma, Clausen:2010hf, Haas:2012bk}, ultra-luminous X-ray binaries \cite{Kaaret:2017tcn}, and gravitational waves (GWs) \cite{Mandel:2007hi, Gair:2010dx, Fragione:2017blf, LIGOScientific:2020iuh}.

However, electromagnetic (EM) observations of IMBHs remain elusive, although some indirect evidence supports their existence \cite{Mezcua:2017npy, Farrell:2009uxm, Pasham2014, Kiziltan2017}. This is primarily due to their typically weak EM emission and the absence of distinct observational features. GW observations, on the other hand, provide a clean and direct probe of black hole masses and dynamics, opening a new window into astronomy \cite{LIGOScientific:2017ync, LIGOScientific:2017zic, LIGOScientific:2017ycc, LIGOScientific:2016vpg, LIGOScientific:2018hze, Safarzadeh:2019pis, LIGOScientific:2017vox, LIGOScientific:2018jsj, Margalit:2017dij, Savchenko:2017ffs}, fundamental physics \cite{DES:2019ccw, DES:2020nay, Nunes:2020rmr, Xu:2023wog, Ezquiaga:2017ekz, Zhu:2022dfq, Li:2023gtu}, and cosmology \cite{LIGOScientific:2018gmd, LIGOScientific:2019zcs, Zhang:2018byx, Du:2018tia, Zhang:2019ylr,  Belgacem:2019tbw, Zhang:2019loq, Zhang:2019ple, He:2019dhl, Caprini:2018mtu, Zhao:2019gyk, Wang:2019tto, Chen:2020dyt, Jin:2020hmc, Borhanian:2020vyr, Chen:2020zoq, Mitra:2020vzq, Bian:2021ini, Ye:2021klk, Jin:2021pcv, Guo2021StandardSC, Yu:2021nvx, deSouza:2021xtg, Dhani:2022ulg, Zhu:2023jti, Hou:2022rvk, Califano:2022syd, Jin:2022qnj, Jin:2023zhi, Han:2023exn, Wang:2023lif, Jin:2023tou, Vagnozzi:2023nrq, DelPozzo:2011vcw, Nair:2018ign, Gray:2019ksv, Zhu:2021aat, Yu:2020vyy, Finke:2021aom, Leandro:2021qlc, Song:2022siz, LIGOScientific:2021aug, Palmese:2021mjm, Gair:2022zsa, Mukherjee:2022afz, Yang:2021xox, Yang:2022tig, Muttoni:2023prw,  Yu:2023ico, Zhu:2024qpp, Xiong:2024gpx, Xiao:2024nmi, Li:2025ula, Li:2025eqh, Han:2025fii, Song:2025ddm, Li:2025owk, Han:2024sxm, Zhang:2024rra,  Li:2024qus, Li:2024qso, Du:2025odq}, and serve as ideal tool for identifying IMBHs and studying their properties. Currently, the LIGO-Virgo-KAGRA GW detector network has detected an IMBH candidate, GW231123 \cite{LIGOScientific:2025rsn}, with a remnant mass of approximately $225\,M_\odot$ and both of its progenitors lying above the pair-instability mass gap \cite{Mehta:2021fgz}. Future ground-based GW detectors such as Einstein Telescope (ET) \cite{Punturo:2010zz} and Cosmic Explorer (CE) \cite{LIGOScientific:2016wof} are expected to reach higher sensitivities, resulting in substantially increased detection rates. Nevertheless, these ground-based GW detectors are primarily sensitive to the high-frequency merger and ringdown phases of IMBH binaries, potentially missing lower-frequency inspiral signals.

To improve the detection of IMBHs, GW observations need to extend into the decihertz frequency band. In recent years, several decihertz-band GW detector projects have been proposed, including space-based missions such as DECi-hertz Interferometer Gravitational Wave Observatory \cite{Kawamura:2006up, Kawamura:2011zz, Seto:2001qf} and Decihertz Observatory \cite{Sedda:2021yhn}, as well as several lunar-based GW detection plans. In fact, compared to space-based GW detectors, lunar-based GW detectors have significant advantages in structural stability and long-term deployment. In addition, the moon has a natural vacuum environment and low seismic noises, which provides an ideal platform for the construction of GW detectors \cite{Lognonne1993, Nakamura1981}. In particular, permanently shadowed regions near the lunar poles maintain extremely low surface temperatures, which help to minimize interference caused by thermal drift \cite{Paige2010}. At present, many concepts of lunar-based detectors have been proposed, including Lunar Gravitational-Wave Antenna (LGWA) \cite{LGWA:2020mma} and Lunar Seismic and Gravitational Antenna (LSGA) \cite{Katsanevas2020LSGA} based on lunar seismometers, as well as laser interferometer plans, such as Gravitational-wave Lunar Observatory for Cosmology (GLOC) \cite{Jani:2020gnz} and Laser Interferometer On the Moon (LION) \cite{Amaro-Seoane:2020ahu}. Among them, LGWA has been the subject of extensive studies, including its scientific objectives and its potential for detecting the GW background and IMBHs \cite{Ajith:2024mie, Cozzumbo:2023gzs, LGWA:2020mma, Yan:2024jik, Song:2025lpa}. We note that the waveform of the inspiral phase of IMBHs mainly falls in the decihertz band, while the waveforms of the merger and ringdown phases fall within the sensitivity band of ground-based GW detectors. Therefore, multi-band detection combining the decihertz and few hertz bands will undoubtedly further enhance the detection of IMBHs.

In recent years, with the proposal of various GW detectors operating in different frequency bands, multi-band detection of GWs is becoming increasingly feasible. Some studies have shown that multi-band detection enables early capture and continuous tracking of GW signals, which improves the signal-to-noise ratio (SNR) and the accuracy of parameter estimation \cite{Sesana:2016ljz,Vitale:2016rfr, Sesana:2017vsj,Isoyama:2018rjb,Jani:2019ffg,Carson:2019kkh,Liu:2020nwz,Datta:2020vcj,Zhang:2021pwe,Nakano:2021bbw,Yang:2021qge,Muttoni:2021veo,Liu:2021dcr,Zhu:2021bpp,Kang:2022nmz,Klein:2022rbf,Seymour:2022teq,Baker:2022eiz,Zhao:2023ilw, Dong:2024bvw}. It also plays an important role in early signal warning, searching for EM counterparts, and accurately locating GW sources. Therefore, in this work, we wish to investigate how multi-band detection by LGWA and ET affects the detection of IMBHs.

We simulate signals from three IMBH population models and carefully explore detector coverage in important parameters like primary mass, mass ratio, inclination angle, and redshift. Our goal is to evaluate how effectively multi-band observations can recover the simulated population distributions from different models. We also aim to assess the benefits of multi-band detection in extending detection range, improving parameter estimation accuracy, and revealing the origins of the source population.

This paper is organized as follows. Section~\ref{Method} outlines the methodology for simulating IMBH merger events in the context of joint LGWA and ET. In Section~\ref{sec:results and discussion}, we present the resulting constraints and discuss their astrophysical implications in detail. Finally, Section~\ref{Conclusion} summarizes the key findings and concludes the study.

\section{Method}\label{Method}

\subsection{Simulation of IMBH source parameters}\label{sec:simulate GW source}

In this study, we adopt a parameterized model to simulate a population of merging IMBH binaries following methods established in the literature~\cite{Fragione:2022ams}. Each binary is characterized by its redshift $z$, primary mass $M_1$, and mass ratio $q$. The volumetric merger rate is given by
\begin{equation}
    \mathcal{R}(z, M_1, q) = K \, \mathcal{N}(\mu_z, \sigma_z) \, M_1^{-\alpha} \, q^{-\beta},
\end{equation}
where $K$ is a normalization constant in units of $\mathrm{Gpc^{-3} \, yr^{-1}}$, and $\mathcal{N}(\mu_z, \sigma_z)$ denotes a Gaussian distribution in redshift with mean $\mu_z$ and standard deviation $\sigma_z$. The parameters $\alpha$ and $\beta$ characterize the power-law distributions of the primary mass $M_1$ and the mass ratio $q$, respectively.

The expected number of merger events per unit redshift, primary mass, and mass ratio is then given by

\begin{equation}
\begin{split}
    \frac{{\rm d}\dot{N}}{{\rm d}z \, {\rm d}M_1 \, {\rm d}q} 
    = \mathcal{R}(z, M_1, q) \, \frac{{\rm d}V_c}{{\rm d}z}
    = K \, \frac{\mathcal{N}(\mu_z, \sigma_z)}{1+z} \, M_1^{-\alpha} \, q^{-\beta},
\end{split}
\end{equation}
where the factor $1/(1+z)$ accounts for cosmological time dilation, and ${\rm d}V_c/{\rm d}z$ is the differential comoving volume element.

To explore different astrophysical scenarios, we consider three representative models, two adopting specific parameterized forms and the other based on uniform distributions. The first model adopts $\{\mu_z, \sigma_z, \alpha, \beta\} = \{2, 1, 1, 1\}$, representing a merger history dominated by repeated hierarchical mergers in dense star clusters~\cite{Fragione:2022egh}. The second model uses $\{5, 1, 1, 1\}$, which corresponds to mergers originating from Population III star remnants and peaking at higher redshift~\cite{Hijikawa:2022qug}. The third model adopts uniform distributions in redshift, primary mass, and mass ratio, providing an agnostic reference scenario independent of astrophysical assumptions.

The remaining binary parameters, which include colatitude $\theta$, longitude $\phi$, inclination angle $\iota$, polarization angle $\psi$, coalescence phase $\varphi_{\rm c}$, and coalescence time $t_{\rm c}$, are sampled uniformly within their physically allowed ranges: $\cos\theta \in [-1,1]$, $\phi \in [0,2\pi)$, $\cos\iota \in [-1,1]$, $\psi \in [0,2\pi)$, $\varphi_{\rm c} \in [0,2\pi)$, and $t_{\rm c} \in [0,10]$ yrs.

This model forms the basis for generating mock IMBH merger populations used in our GW simulations.
In our simulations, all calculations involving the conversion between redshift and distance adopt a flat $\Lambda$CDM cosmology, with a Hubble constant of $H_0 = 67.27\ \mathrm{km\,s^{-1}\, Mpc^{-1}}$ and a present-day matter density parameter of $\Omega_{\rm m} = 0.3166$, consistent with the Planck 2018 results~\cite{Planck:2018vyg}.

\subsection{Simulation of GW signals}

The frequency-domain waveform vector for a network of $N$ detectors is given by~\cite{Wen:2010cr,Zhao:2017cbb}
\begin{equation}
\bm{\tilde{h}}(f) = e^{-i\Phi} \bm{\hat{h}}(f),
\end{equation}
where $\Phi$ is an $N \times N$ diagonal matrix with elements $\Phi_{kl} = 2\pi f \delta_{kl} (\bm{n} \cdot \bm{r}_k)$, with $\bm{n}$ the GW propagation direction and $\bm{r}_k$ the position of the $k$-th detector. The waveform vector $\bm{\hat{h}}(f)$ has components
\begin{equation}
\bm{\hat{h}}(f) = \left[\tilde{h}_1(f), \tilde{h}_2(f), \cdots, \tilde{h}_N(f)\right],
\end{equation}
with each $\tilde{h}_k(f)$ given by
\begin{equation}
\tilde{h}_k(f) = h_+(f) F_{+,k}(f) + h_\times(f) F_{\times,k}(f),
\end{equation}
where $h_+(f)$ and $h_\times(f)$ are the two GW polarizations, and $F_{+,k}$ and $F_{\times,k}$ are the antenna pattern functions.

In this work, We employ the \texttt{GWFish}\footnote{\url{https://github.com/janosch314/GWFish}} package~\cite{DupletsaHarms2023} together with the inspiral merger ringdown (IMR) waveform model IMRPhenomD~\cite{Husa:2015iqa,Khan:2015jqa} to simulate GW signals. The analysis uses ET and LGWA, which represent ground-based and lunar-based detectors, respectively.

ET consists of three 10-km interferometers forming an equilateral triangle. The antenna pattern for one interferometer is given by~\cite{Punturo:2010zz, Hild:2010id, Zhao:2010sz, ETsensitivities}
\begin{equation}
\begin{split}
\begin{aligned}
F_{+}(\theta,\phi,\psi) &= \frac{\sqrt{3}}{2} \Big[ \frac{1}{2}(1+\cos^2\theta)\cos2\phi\cos2\psi
 - \cos\theta\sin2\phi\sin2\psi \Big], \\
F_{\times}(\theta,\phi,\psi) &= \frac{\sqrt{3}}{2} \Big[ \frac{1}{2}(1+\cos^2\theta)\cos2\phi\sin2\psi
 + \cos\theta\sin2\phi\cos2\psi \Big].
\end{aligned}
\end{split}
\end{equation}
The other two interferometers have pattern functions shifted by $2\pi/3$ and $4\pi/3$ in azimuthal angle $\phi$.

For LGWA, we adopt a simplified response model implemented in \texttt{GWFish}. The detector is fixed on the lunar surface, with orientation defined by geographic coordinates and azimuth. The model accounts for the Moon's rotation and orbital motion but neglects internal lunar dynamics. LGWA has two proposed configurations employing either niobium or silicon as the proof mass and suspension material \cite{Ajith:2024mie}. In this work, we focus on the silicon-based configuration with better sensitivity. This simplified model is appropriate for preliminary sensitivity and parameter estimation studies.

\begin{figure}[!htbp]
\includegraphics[width=0.8\textwidth]{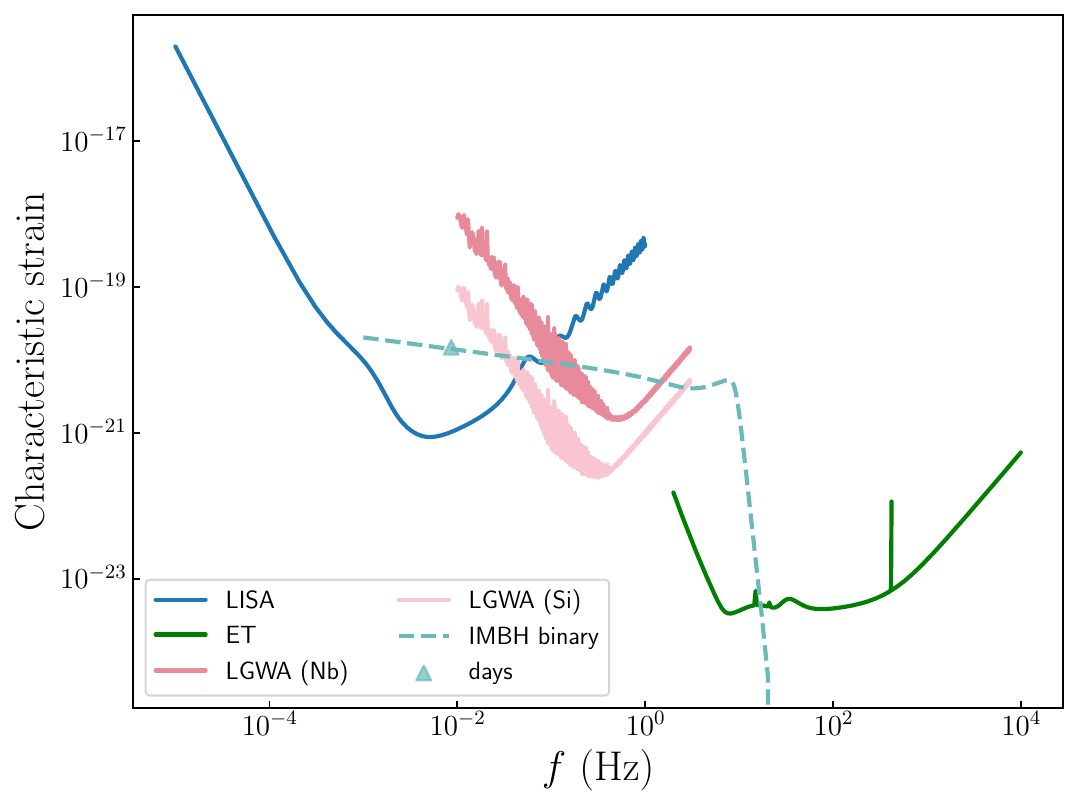}
\centering
\caption{Characteristic strains of ET, LISA, and LGWA (two configurations), along with that of a merging IMBH binary with component masses of $1000$--$1000\,M_\odot$ at redshift $z=1$. The soft cyan triangle marks the frequency one day before coalescence. Characteristic strains are defined as $\sqrt{fS_{\rm n}}$ for the sensitive curves of detectors and $2f|h(f)|$ for the GW signal.}
\label{fig1}
\end{figure}

Figure~\ref{fig1} shows the characteristic sensitivities of ET, LISA, and LGWA in two configurations, along with the simulated GW signal from an IMBH binary merger. The LGWA sensitivity band effectively bridges the gap between space-based and ground-based detectors, enabling continuous multi-band GW observation and accumulation of SNR across a wide frequency range.

\subsection{Calculation of SNR}\label{SNR}

We adopt an SNR detection threshold of 8 for both individual detectors and detector networks, consistent with previous GW studies~\cite{LIGOScientific:2014pky,LIGOScientific:2017adf,LIGOScientific:2016fbo}.

For a network of $N$ detectors, the total SNR $\rho$ is computed as
\begin{equation}
    \rho = \sqrt{\sum_{k=1}^{N} \left( \tilde{h}_k | \tilde{h}_k \right)},
\end{equation}
where the inner product is defined by
\begin{equation}
    \left( \tilde{h} | \tilde{h} \right) = 4 \int_{f_{\rm in}}^{f_{\rm out}} \frac{\tilde{h}(f) \tilde{h}^*(f)}{S_{\rm n}(f)} \, df,
\end{equation}
where $S_{\rm n}(f)$ denotes the one-sided power spectral density (PSD). Here, we adopt the PSDs for LGWA and ET from~\cite{LGWA:2020mma} and~\cite{Hild:2010id}, respectively. Additionally, $f_{\rm in}$ and $f_{\rm out}$ denote the frequencies at which the GW signal enters and exits the detector's sensitive frequency band.

\subsection{Calculation of the detectable population}\label{sec:FIM}

In the following, we describe the method used to construct the detector's detectable population. For a set of $N$ GW events, the corresponding detectable population can be written as
\begin{equation}
\tilde{p}_{\rm det}(\theta_{\rm det})
= 
\frac{1}{N}
\sum_{i=1}^{N}
P_{\rm det}(\theta_i)\,
K(\theta_{\rm det}\mid\theta_i),
\end{equation}
where $P_{\rm det}(\theta_i)$ is the probability that the event $\theta_i$ is detected, which depends on the SNR. $K(\theta_{\rm det} \mid \theta_i)$ represents the conditional probability of observing the parameter $\theta_{\rm det}$ given the true value $\theta_i$, and is expressed as
\begin{equation}
K(\theta_{\rm det} \mid \theta_i) 
= \frac{1}{\sqrt{2\pi \, \sigma_i^2}} 
\exp\Big[-\frac{(\theta_{\rm det}-\theta_i)^2}{2\sigma_i^2}\Big],
\end{equation}
where $\sigma_i$ is the standard deviation of the parameter, calculated from the Fisher information matrix (FIM) \cite{Finn:1992wt}. We use the publicly available package \texttt{GWFish} to estimate LGWA's capability to measure parameters of IMBH via the FIM method.

For a network consisting of $N$ interferometers, the FIM is defined as
\begin{equation}
F_{ij} = \sum_{k=1}^{N} 
\left( 
\frac{\partial \tilde{h}_{k}}{\partial \theta_i} 
\Bigg| 
\frac{\partial \tilde{h}_{k}}{\partial \theta_j} 
\right),
\end{equation}
where $\tilde{h}_k$ denotes the GW waveform observed by the $k$th detector, and $\theta_i$ is the $i$th element of the parameter vector
\[
\boldsymbol{\theta} = 
\{ d_{\rm L},\, t_{\rm c},\, \mathcal{M}_{\rm c},\, \eta,\, 
   \theta,\, \phi,\, \psi,\, \iota,\, \psi_{\rm c} \}.
\]
These parameters describe the GW signal, including the luminosity distance $d_{\rm L}$, coalescence time $t_{\rm c}$, chirp mass $\mathcal{M}_{\rm c}$, symmetric mass ratio $\eta$, sky position angles $(\theta, \phi)$, polarization angle $\psi$, inclination angle $\iota$, and coalescence phase $\psi_{\rm c}$.

The $9\times9$ covariance matrix of the source parameters is obtained by inverting the FIM. The $1\sigma$ uncertainty for each parameter is then given by $\Delta\theta_i = \sqrt{\mathrm{Cov}_{ii}}$.

\section{Results and discussion}\label{sec:results and discussion}

In this section, we first present the detection horizons of three detectors sensitive to different frequency bands, highlighting the frequency advantage of LGWA. We then examine how key source parameters such as primary mass, mass ratio, redshift, and inclination angle affect the detection rate. Finally, we compare the performance of different GW detectors in recovering the intrinsic source population.

\begin{figure}[!htbp]
\includegraphics[width=0.8\textwidth]{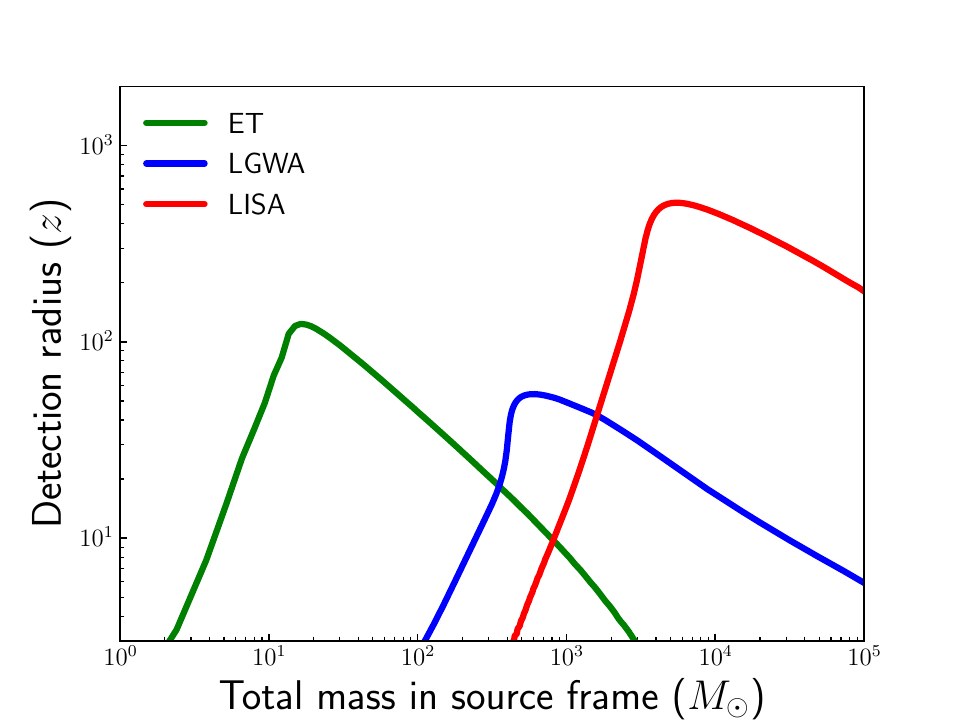}
\centering
\caption{Detection horizons for equal-mass, non-spinning binary black holes as a function of total source-frame mass for ET, LGWA, and LISA detectors.}
\label{fig:horizen}
\end{figure}

Figure~\ref{fig:horizen} illustrates the detection range of three types of GW detectors operating in different frequency bands for binary black hole systems. Compared to the multi-band observation with LISA and ET, the LGWA+ET combination covers a larger detection horizon, enabling the observation of binary black hole systems across a wider mass range and extending to higher redshifts.
As shown in Figure~\ref{fig1}, this advantage is primarily due to LGWA's sensitivity in the decihertz band, which lies closer in frequency to the operating range of ET than LISA does. This relative frequency proximity facilitates more continuous spectral coverage, thereby improving the detectability of IMBH binary mergers in multi-band observations.

\begin{figure*}[htbp]
    \centering
    \includegraphics[width=0.32\textwidth]{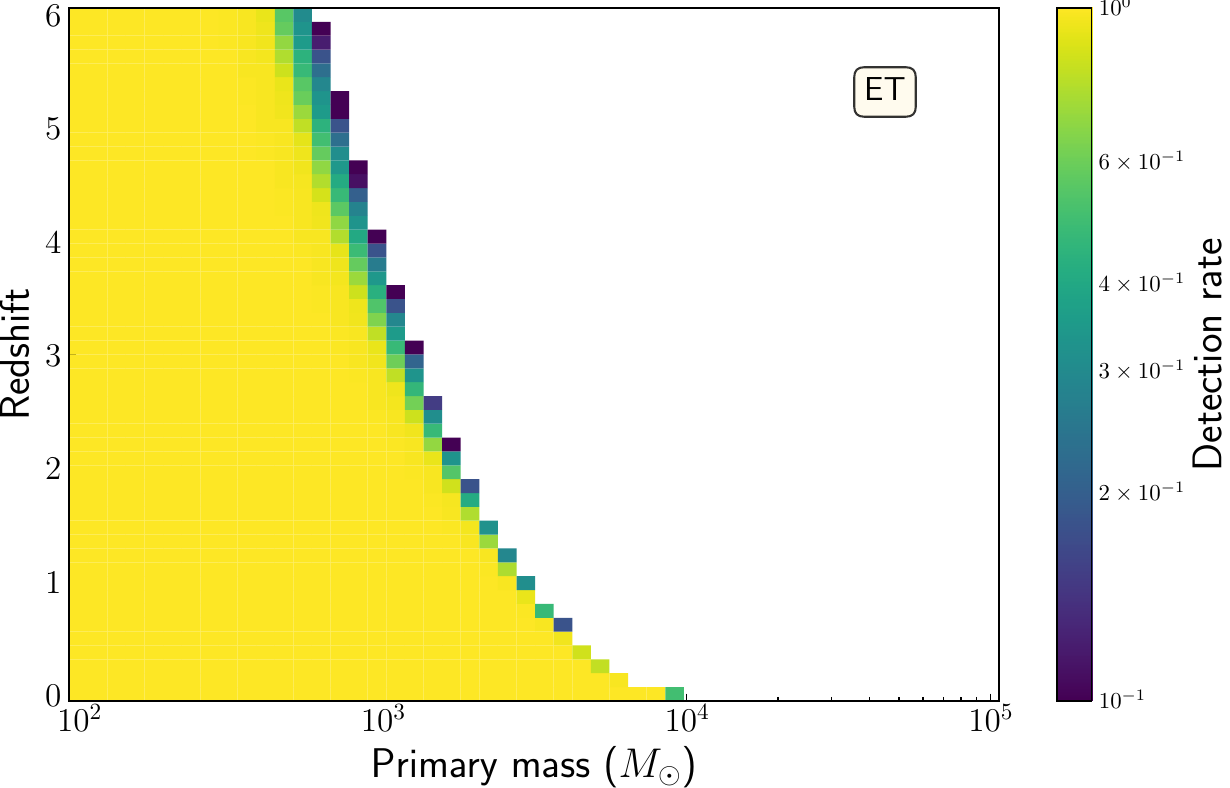}
    \includegraphics[width=0.32\textwidth]{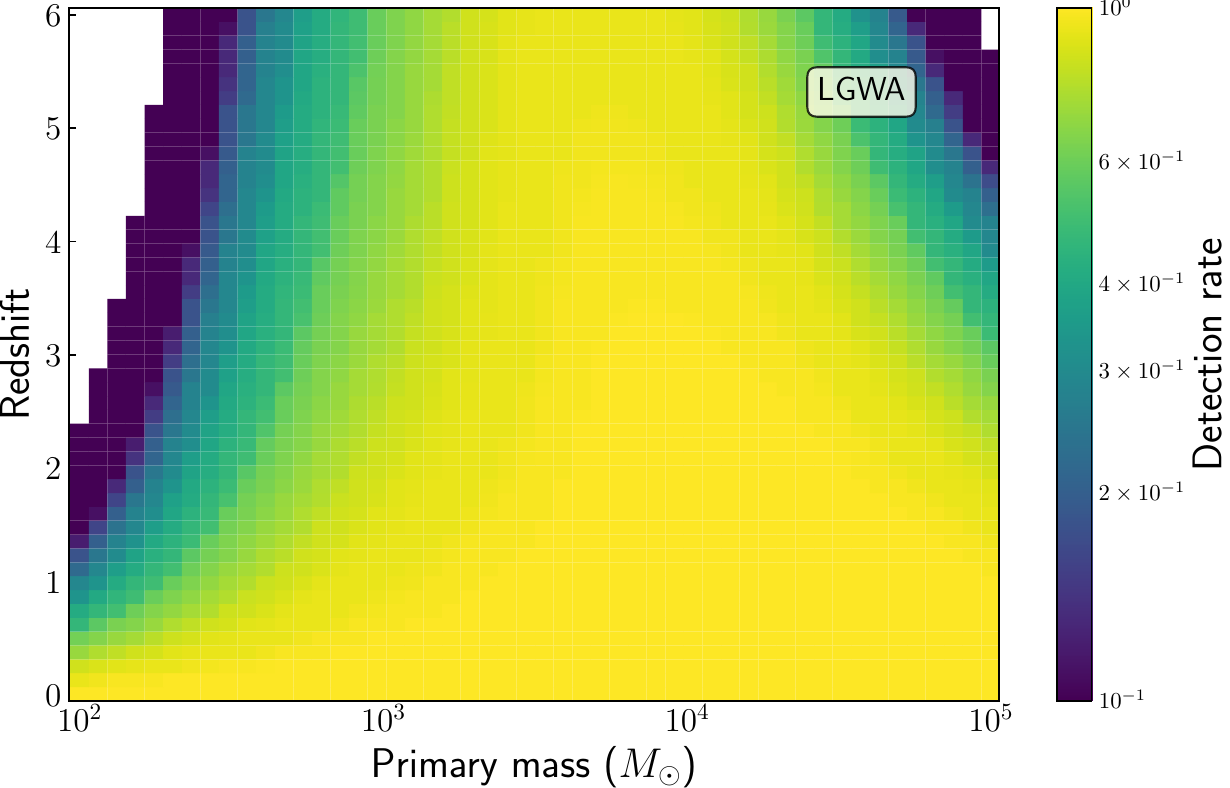}
    \includegraphics[width=0.32\textwidth]{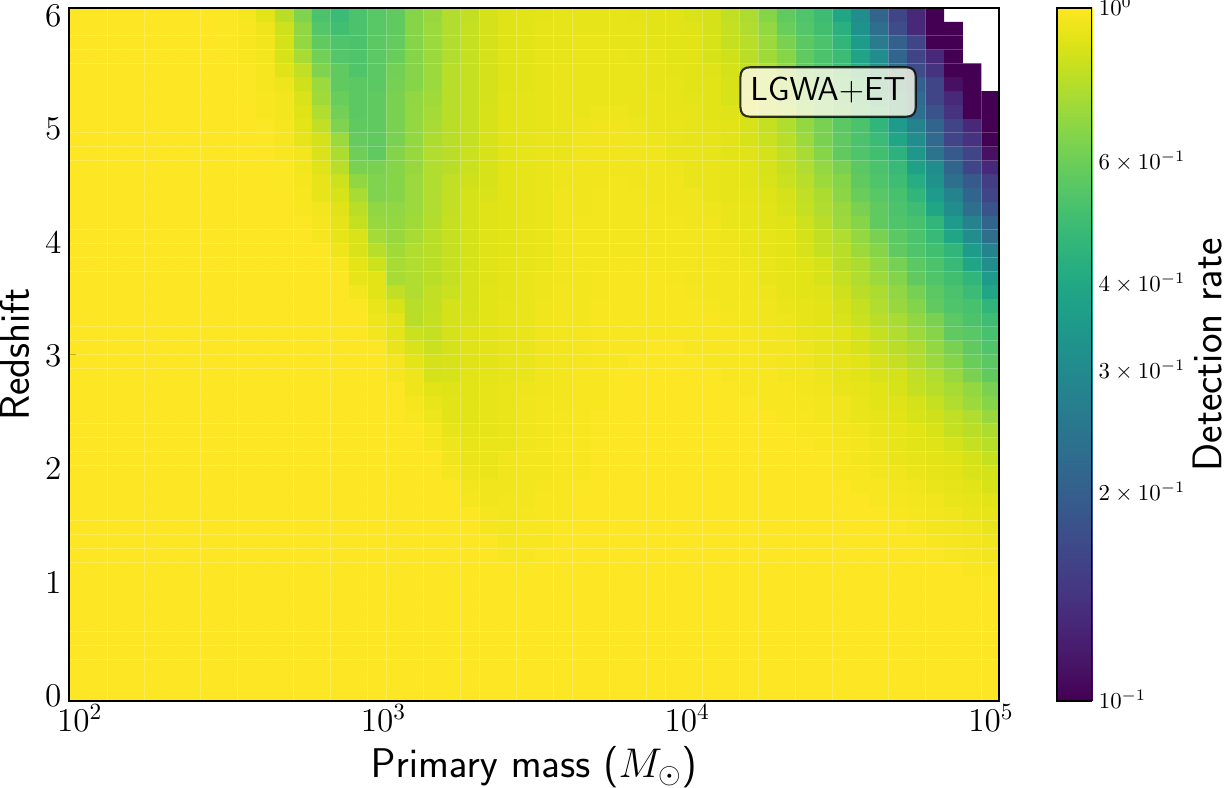}
    \caption{Detection rate of binary black hole mergers as a function of the primary mass $M_1$ and redshift $z$, for different GW detector configurations. The color scale indicates the relative detection rate.
    \textbf{Top:} ET only.
    \textbf{Middle:} LGWA only.
    \textbf{Bottom:} Multi-band detection with both ET and LGWA.}
    \label{fig:z_m}
\end{figure*}

\begin{figure*}[htbp]
    \centering
    \includegraphics[width=0.32\textwidth]{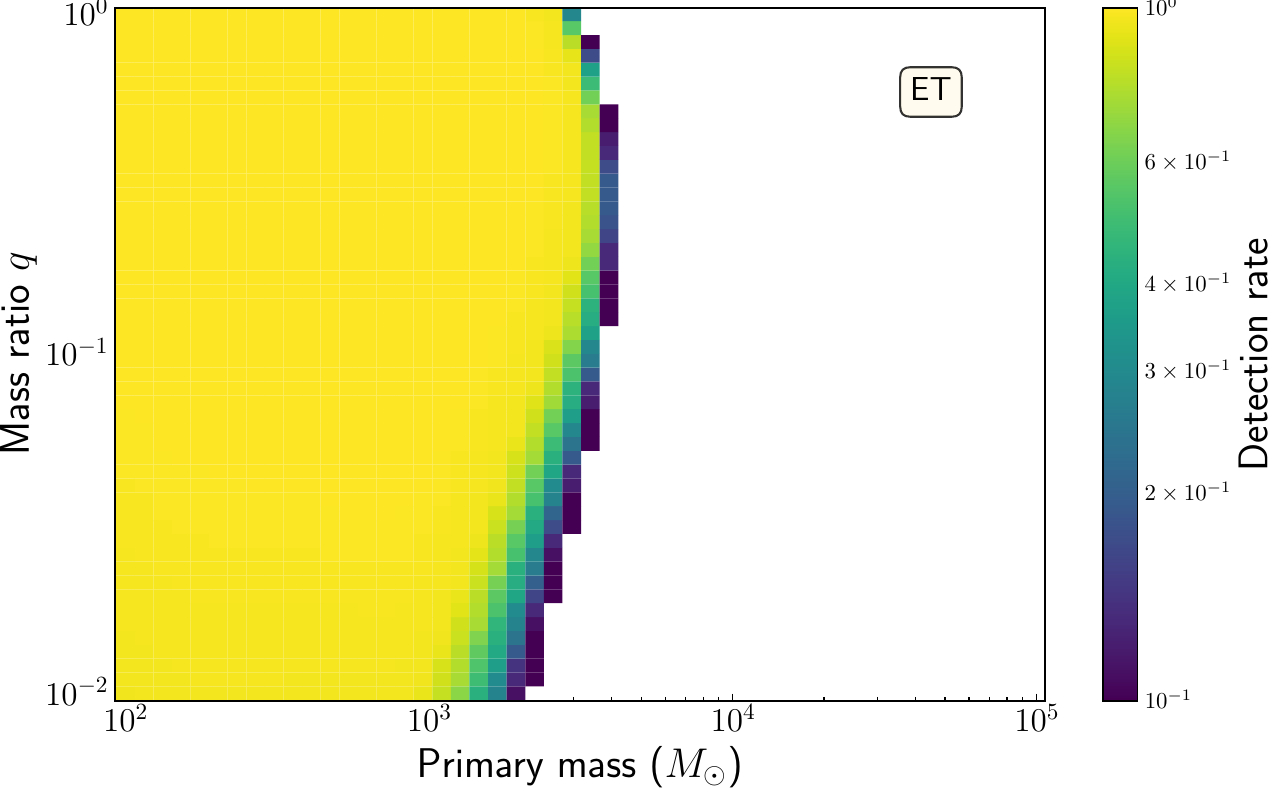} 
    \includegraphics[width=0.32\textwidth]{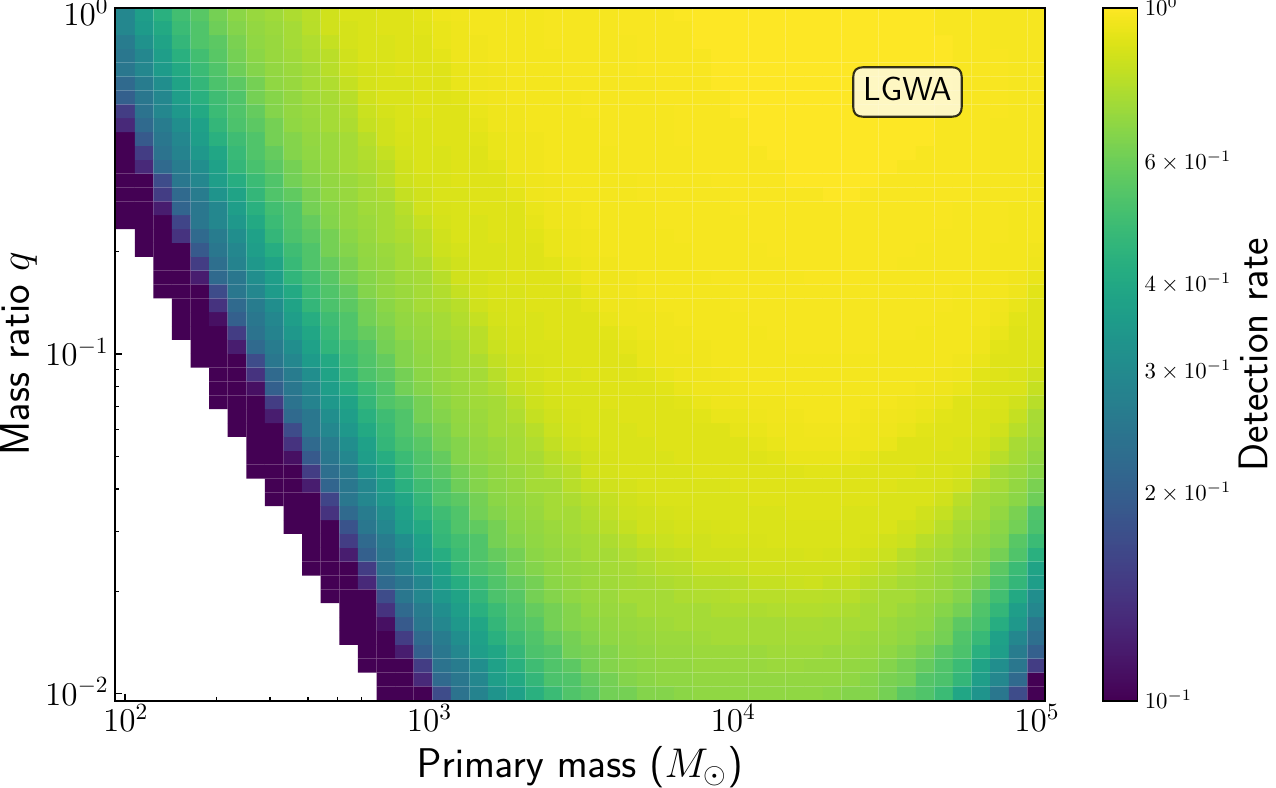} 
    \includegraphics[width=0.32\textwidth]{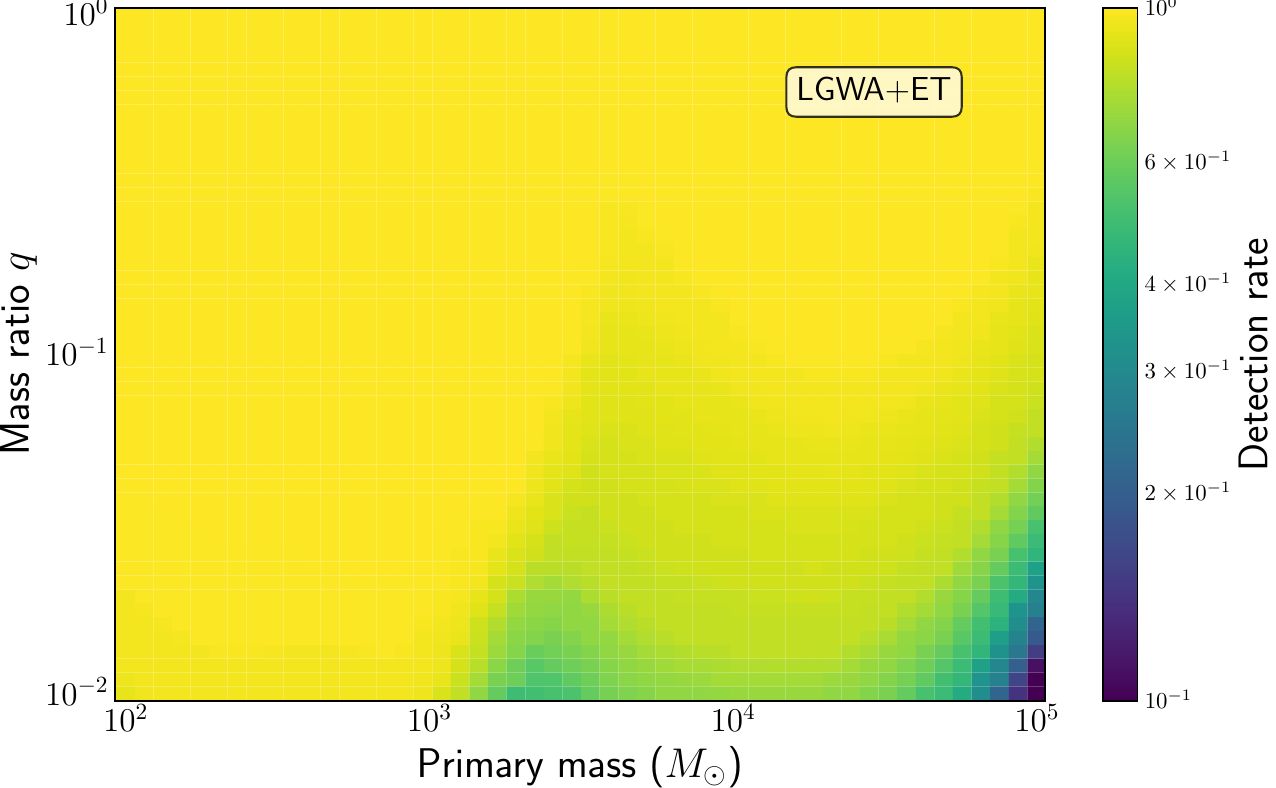}
    \caption{Similar to Figure~\ref{fig:z_m}, showing the detection rate of binary black hole mergers at redshift $z=1$, but in the parameter space of primary mass $M_1$ and mass ratio $q$.
    }
    \label{fig:q_m}
\end{figure*}

\begin{figure*}[htbp!]
    \centering
    \includegraphics[width=0.32\textwidth]{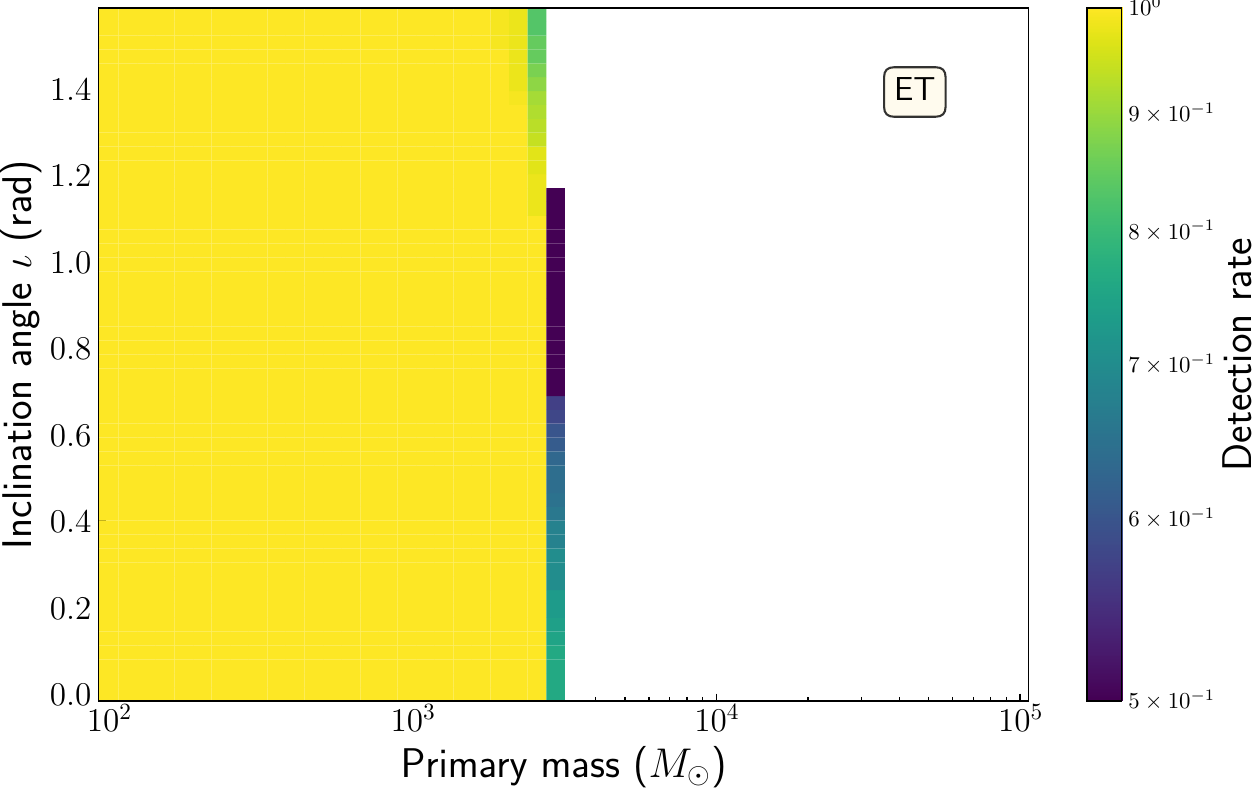} 
    \includegraphics[width=0.32\textwidth]{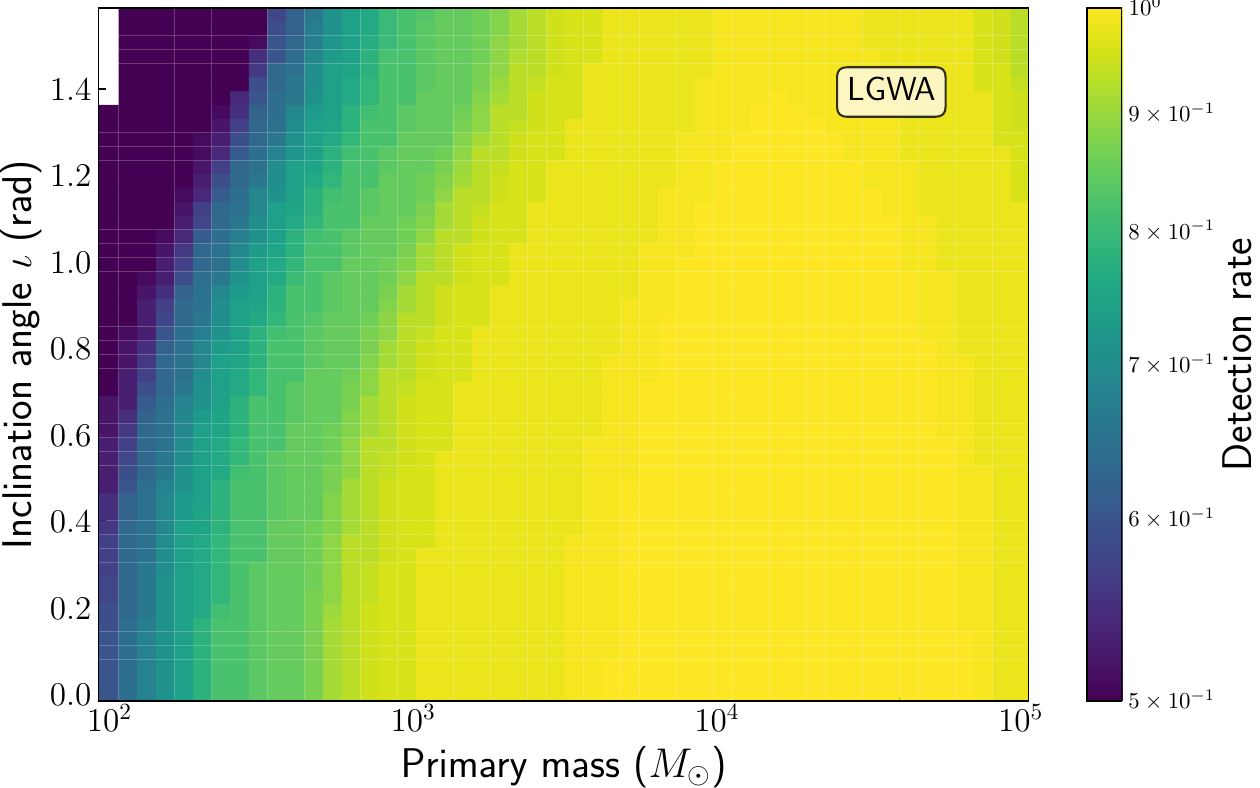} 
    \includegraphics[width=0.32\textwidth]{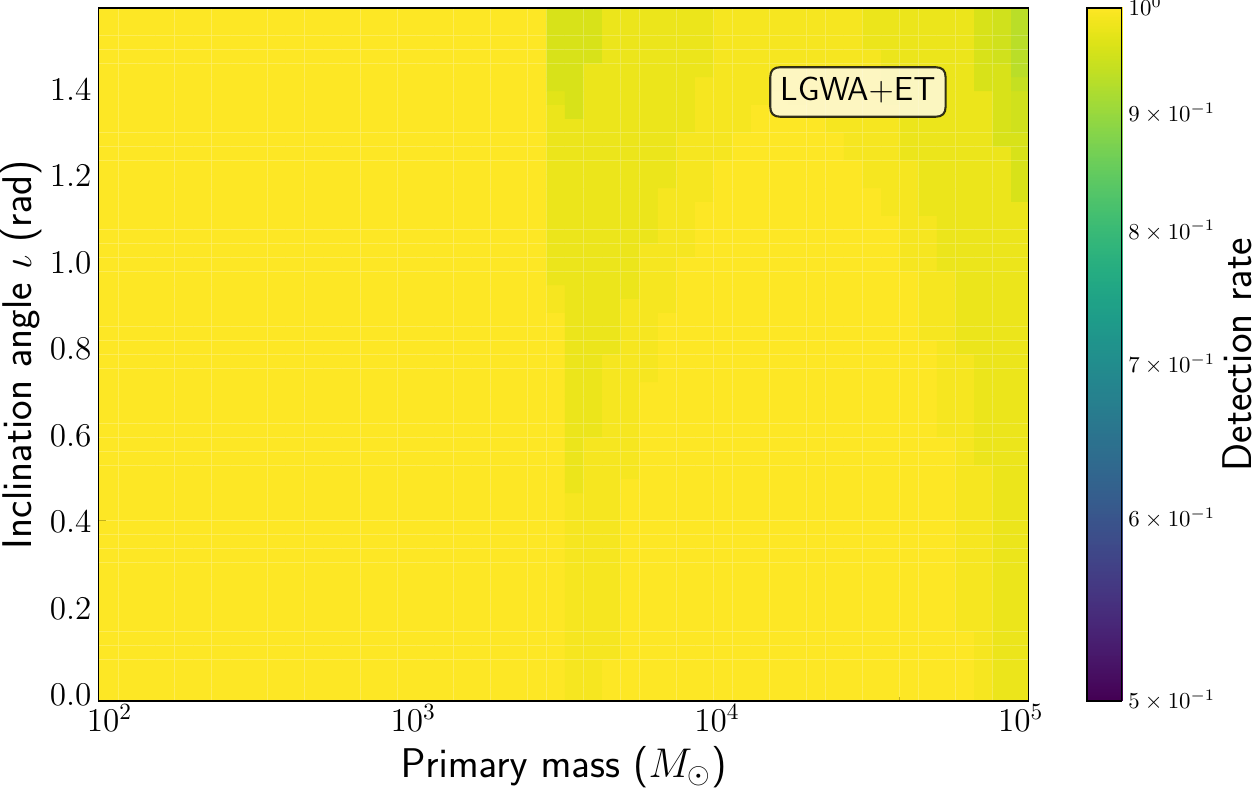}
    \caption{Similar to Figures~\ref{fig:z_m} and \ref{fig:q_m}, showing the detection rate in the parameter space of primary mass $M_1$ and inclination angle $\iota$ at redshift $z = 1$.}
    \label{fig:iota_m}
\end{figure*}

To clearly demonstrate the advantages of multi-band observations by LGWA and ET in detecting IMBHs, we present the detection abilities of different GW detectors across the redshift-primary mass, mass ratio-primary mass, and inclination angle-primary mass parameter spaces, as shown respectively in Figures~\ref{fig:z_m}--\ref{fig:iota_m}.

In Figure~\ref{fig:z_m}, we present the detection rate of binary black hole mergers across the parameter space of primary mass $M_1$ and redshift $z$ for three different detector configurations. Here, the detection rate represents the fraction of IMBH binaries that can be detected at each point in the $(M_1, z)$ plane, where an event is considered detectable if its SNR exceeds the threshold of 8. The redshift range is chosen based on the model parameters \(\{\mu_z, \sigma_z, \alpha, \beta\} = \{2, 1, 1, 1\}\), with an upper limit of \(z = 6\).
The top panel shows the detection performance of ET. Since ET is primarily sensitive to frequencies above 10 Hz, its detection capability is concentrated in the region of low masses and low redshifts. At low redshifts, ET can detect primary masses up to approximately $10^{4} M_{\odot}$, but as redshift increases to around 6, the detectable range decreases to about $800 M_{\odot}$. As the primary mass and redshift increase, the GW signals remain within ET's sensitive frequency band for a shorter duration, resulting in a rapid decline in the detection rate. The middle panel corresponds to LGWA alone. With sensitivity in the decihertz frequency band, LGWA can detect more massive black hole binaries, achieving a high detection rate over the range $M_1 \sim 10^3$--$10^5\,M_\odot$ and extending to redshifts beyond $z \sim 6$. Notably, the detection rates for both high-mass and low-mass binaries decrease as redshift increases, while the detection rate remains relatively high in the intermediate mass range. 
The bottom panel presents the combined performance of ET and LGWA. Within the considered $M_1$-$z$ parameter space, the multi-band observation involving multiple detectors achieves nearly complete coverage. This improvement results from the complementary sensitivity of the two detectors, with ET being more effective at detecting lower-mass IMBH mergers and LGWA extending the reach toward higher-mass systems. Although detection capability remains limited at the extremely high-mass or high-redshift ends, the overall detection rate is markedly improved, which further emphasizes the advantages of multi-band GW observations in extending the mass coverage and enhancing detection capabilities at high redshifts.

We further examine how the mass ratio $q$ influences the detectability of IMBH mergers, as illustrated in Figure~\ref{fig:q_m}.
The top panel shows the result for ET alone. Due to its limited sensitivity at low frequencies, ET primarily detects relatively low-mass compact binaries. When the primary mass is below $10^{3}\,M_{\odot}$, ET can detect these sources with a high detection rate, and the mass ratio has little effect on the detection. However, within the total mass range of $[1 \times 10^{3},\, 5 \times 10^{3}]\,M_{\odot}$, the detection rate depends on the mass ratio, with highly asymmetric binaries falling outside the sensitive band and becoming essentially undetectable.
The middle panel corresponds to LGWA, which is particularly sensitive to binary black hole systems with masses in the range $[5 \times 10^3,\, 1 \times 10^5]\,M_\odot$ and mass ratios greater than 0.1. As the total mass decreases or the mass ratio decreases, the signal strength weakens, and the detection rate decreases. 
The bottom panel shows ET and LGWA combined. The multi-band observation provides extensive coverage across the mass and mass-ratio parameter space. Although detection sensitivity decreases for systems with high primary mass and highly asymmetric mass ratios, ET effectively complements LGWA by covering the low-mass and asymmetric regions, significantly enhancing overall detection capabilities.
This highlights the clear advantage of multi-band GW observations in extending mass coverage and improving sensitivity to asymmetric mass-ratio systems.

To investigate the effect of orbital inclination $\iota$ on detection, we present the detection rate in the parameter space of primary mass $M_1$ and inclination angle $\iota$ in Figure~\ref{fig:iota_m}.
Regarding the top panel, which represents the detection performance of ET alone, the detection rate primarily depends on the total mass of the IMBH binary, with only a slight decrease observed at larger inclination angles near the edges of ET's sensitive mass range. This is mainly because ET's high sensitivity and triangular design allow it to effectively measure both polarizations, thereby reducing dependence on the source's inclination angle.
In contrast, the detection rate shown in the middle panel for LGWA is significantly affected by the inclination angle.
Benefiting from its excellent sensitivity in the decihertz band, LGWA exhibits strong detection capability in the high-mass regime. In the mass range of $10^3$ to $10^5\,M_{\odot}$, the detector noise is low and the SNR is high, so the detection rate shows little dependence on the amplitude, and no significant correlation with the inclination angle $\iota$ is observed.
However, in the low-mass regime, as LGWA's noise increases and its sensitivity decreases, the SNR reduces, causing the detection rate to be more significantly affected by the amplitude and correspondingly more sensitive to the inclination angle $\iota$.
The bottom panel presents the combined performance of ET and LGWA. ET's strong sensitivity in the low-mass regime effectively complements LGWA's limitations in this region, leading to effective detection coverage across the entire parameter space of primary mass and inclination angle. 
This demonstrates the advantage of multi-band observations in reducing inclination selection effects and enhancing detection capabilities, significantly decreasing the dependence on the inclination angle $\iota$ across the full mass range of IMBHs.

\begin{figure*}[htbp]
    \centering
    \includegraphics[width=0.32\textwidth]{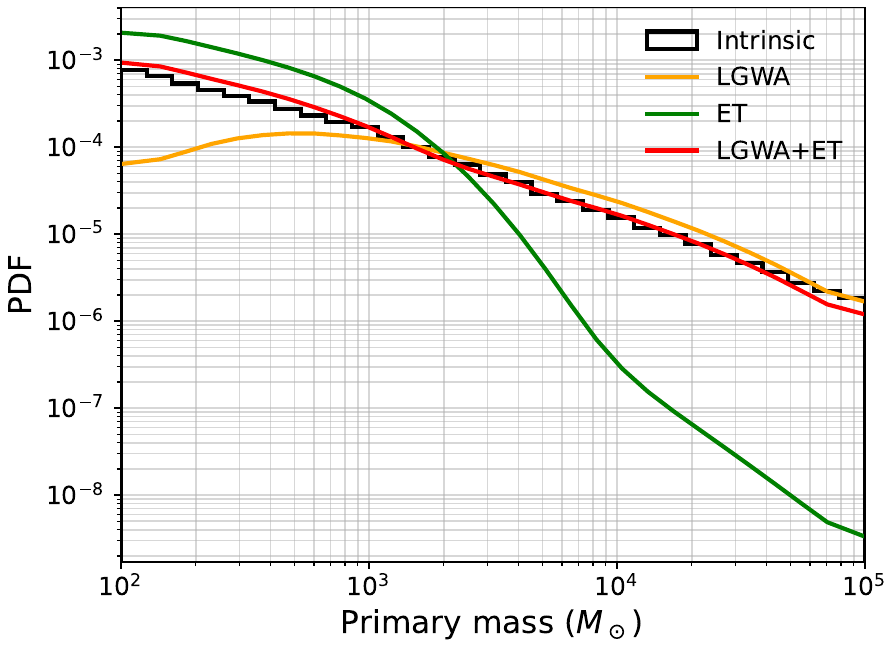}
    \includegraphics[width=0.32\textwidth]{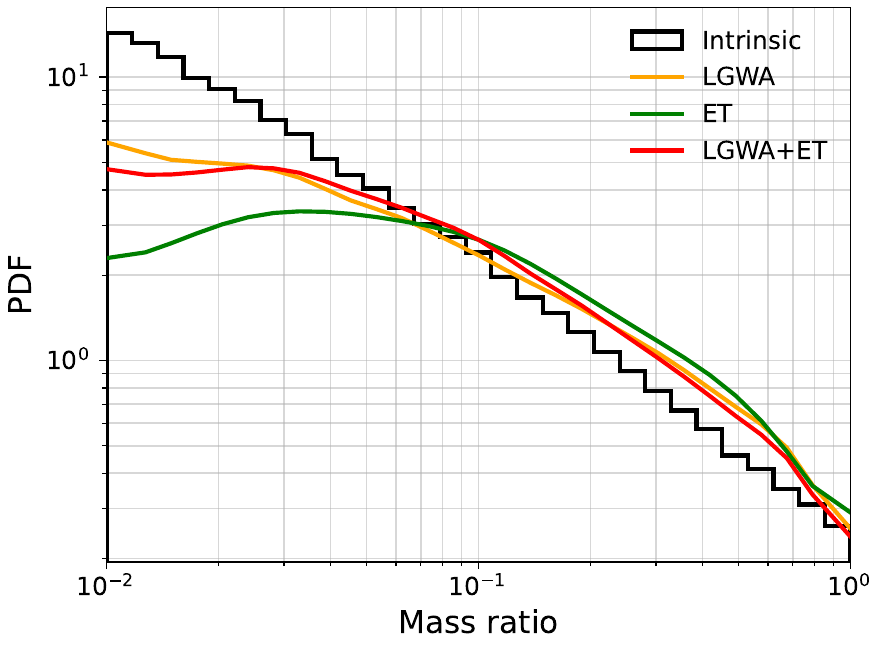}
    \includegraphics[width=0.32\textwidth]{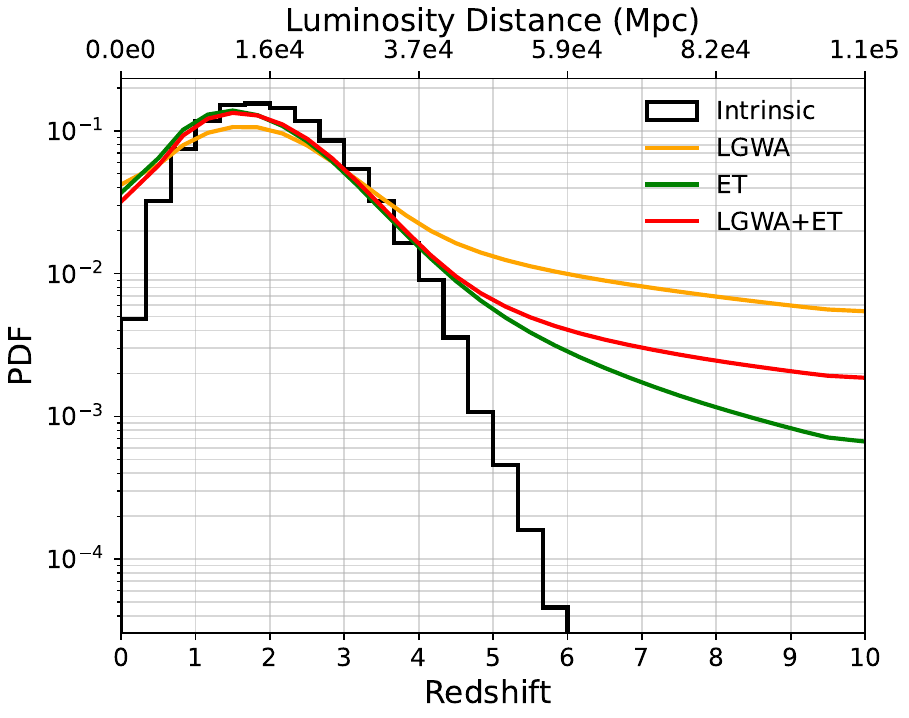}
    \caption{Probability distribution functions (PDFs) of the detected primary mass (left), mass ratio (center), and redshift (right) of merging IMBH binaries, assuming an intrinsic population model with $\{\mu_z, \sigma_z, \alpha, \beta\} = \{2, 1, 1, 1\}$. {The black lines represent the intrinsic distribution, while the green, orange, and red lines correspond to detectable populations for ET, LGWA, and the combined LGWA+ET network, respectively. The detectable populations are obtained by constructing, for each event with a SNR above eight, a Gaussian posterior based on its FIM and combining these posteriors to form the final observable distribution (see Sec.~\ref{sec:FIM} for details).}}
    \label{fig:2111}
\end{figure*}

\begin{figure*}[htbp]
    \centering
    \includegraphics[width=0.32\textwidth]{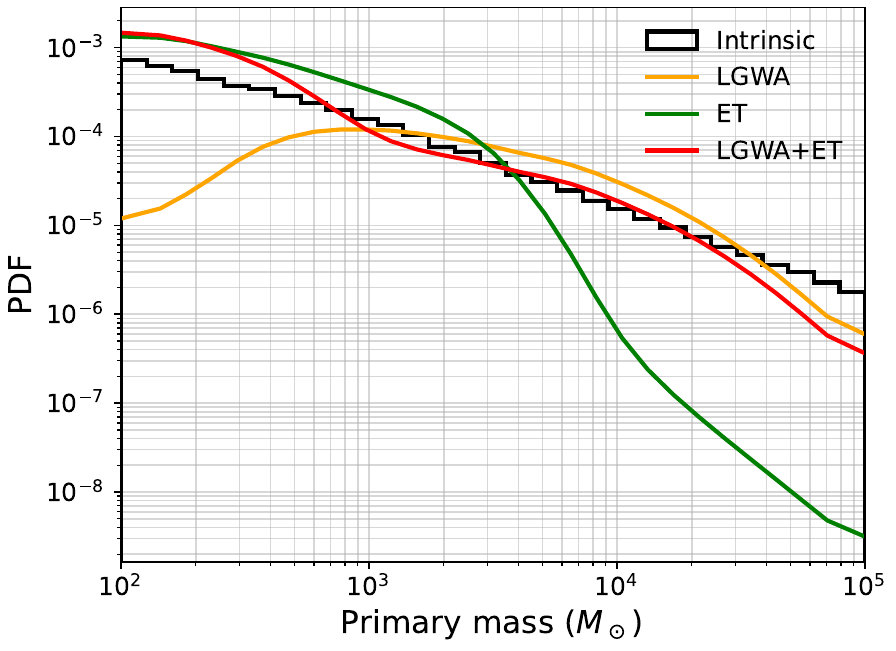}
    \includegraphics[width=0.32\textwidth]{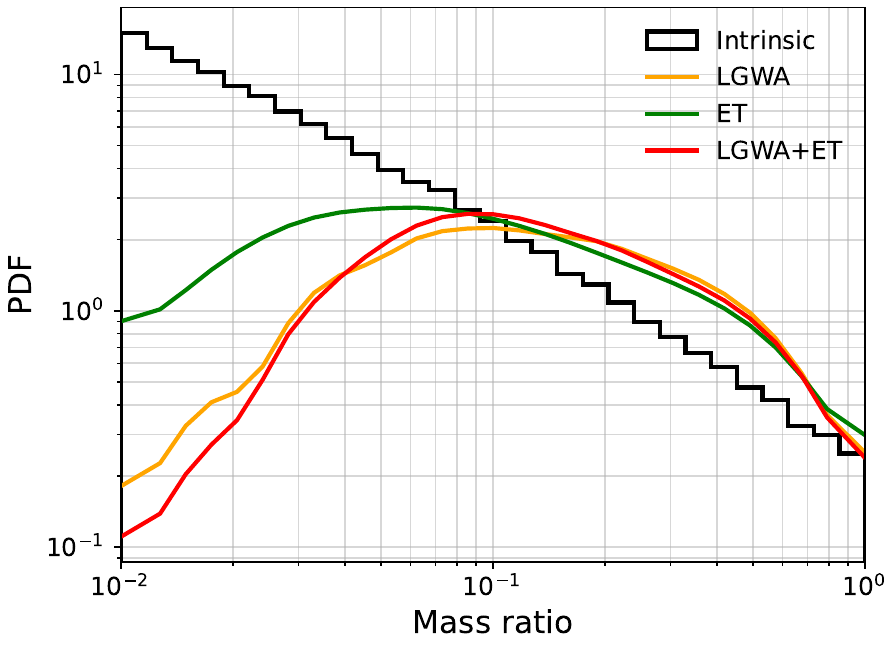}
    \includegraphics[width=0.32\textwidth]{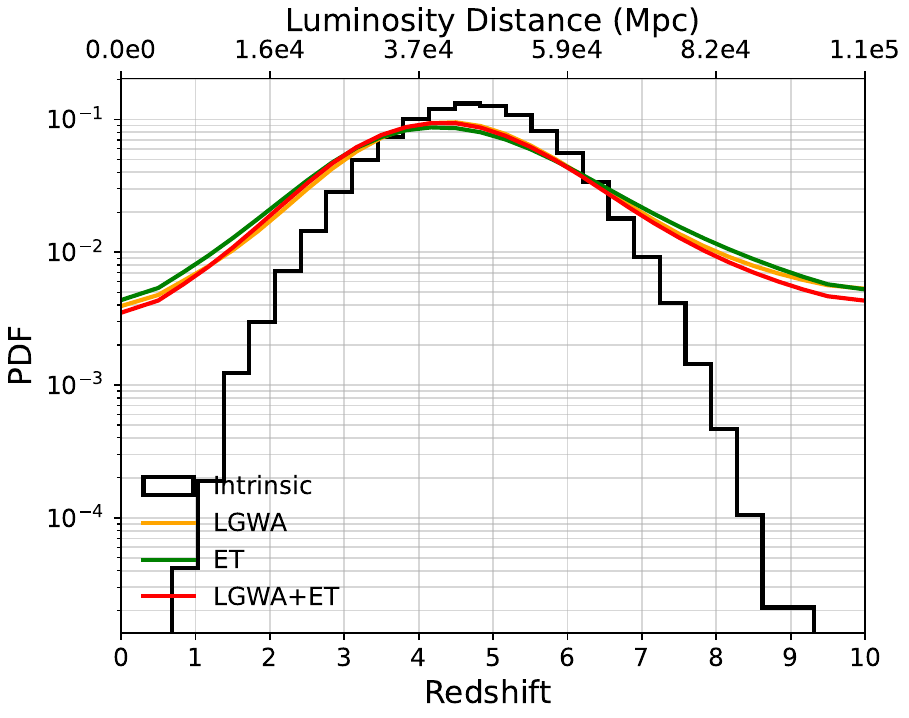}
    \caption{Same as Figure~\ref{fig:2111}, but assuming an intrinsic population model with $\{\mu_z, \sigma_z, \alpha, \beta\} = \{5, 1, 1, 1\}$.}
    \label{fig:5111}
\end{figure*}

\begin{figure*}[htbp]
    \centering
    \includegraphics[width=0.32\textwidth]{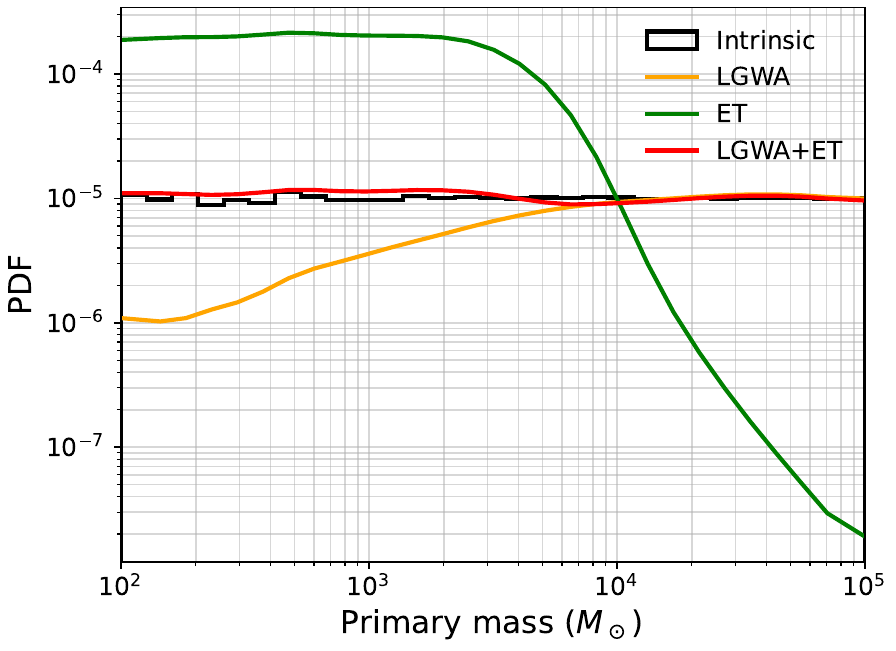}
    \includegraphics[width=0.32\textwidth]{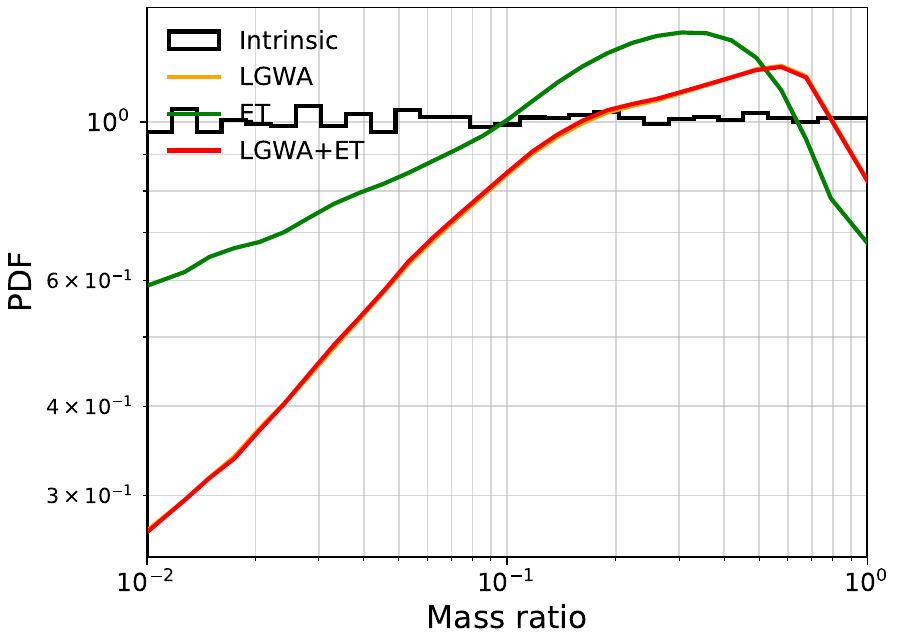}
    \includegraphics[width=0.32\textwidth]{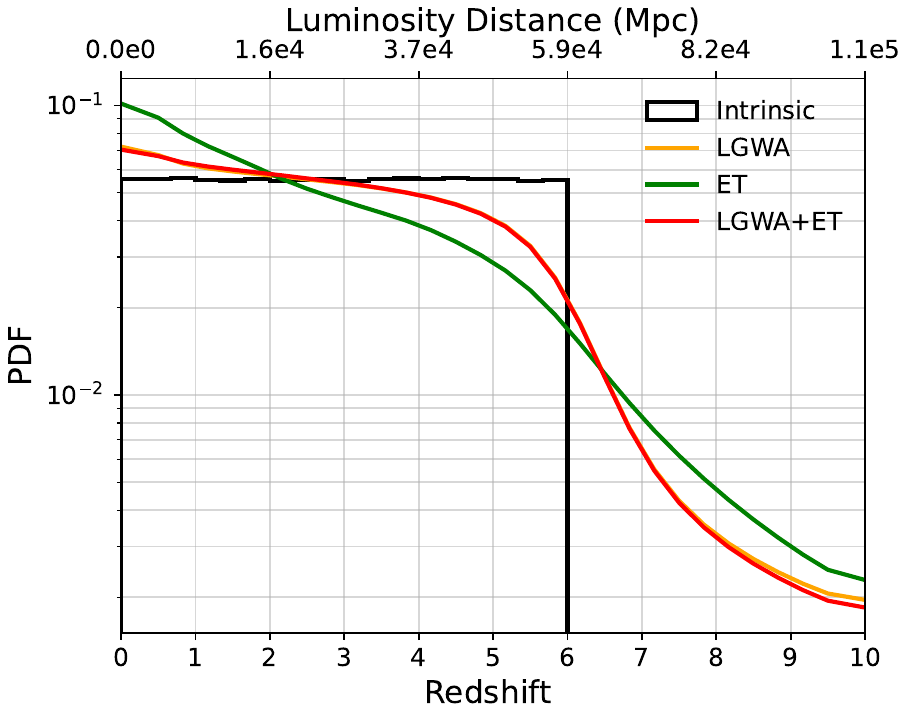}
    \caption{Same as Figures~\ref{fig:2111} and \ref{fig:5111}, but assuming a population model with uniform distributions in redshift, primary mass, and mass ratio, serving as a non-parametric astrophysical baseline.}
    \label{fig:uniform}
\end{figure*}

To explore the advantages of multi-band observations for detecting IMBH population distributions, in Figures~\ref{fig:2111}--\ref{fig:uniform}, we compare the detection capabilities of ET, LGWA, and their combination in three different IMBH population models. Each figure presents the normalized probability density functions (PDFs) of the primary mass $M_1$, mass ratio $q$, and redshift $z$ in three separate panels. \blue{}

The left panels of Figures~\ref{fig:2111}--\ref{fig:uniform} show the distributions of primary masses. We find that ET alone exhibits limited ability to detect high-mass GW events. As shown in Figure~\ref{fig:2111}, the detectable distribution of ET begins to significantly deviate from the intrinsic distribution based on GW population models when the primary mass exceeds approximately $10^3\,M_\odot$. Compared with the scenario in which the IMBH redshift distribution is modeled as a Gaussian, assuming a uniform redshift distribution allows ET to better recover the intrinsic mass distribution. As shown in Figure~\ref{fig:uniform}, under this assumption the mass scale at which ET's detectable distribution begins to deviate substantially from the intrinsic one is pushed to $\gtrsim 10^4\,M_\odot$. This is because a uniform redshift distribution yields more low-redshift IMBH events, and the increased fraction of such detections extends the overall mass range accessible to ET. In comparison, LGWA exhibits limited ability to recover the primary-mass distribution for systems with $M_1 < 10^3\,M_\odot$ across all three models. Its sensitivity to high-mass systems also decreases at higher redshifts. When LGWA is combined with ET, the joint network substantially improves the recovery of the intrinsic primary-mass distribution across the full parameter space, particularly in Figures~\ref{fig:2111} and \ref{fig:uniform}. However, in Figure~\ref{fig:5111}, the higher redshift of the population leads to persistent discrepancies between the detected and intrinsic distributions in both the low-mass ($<10^3\,M_\odot$) and high-mass ($>10^4\,M_\odot$) regimes.

In terms of mass ratio $q$ (middle panels), Figure~\ref{fig:2111} illustrates that ET has limited capability to recover the distribution of highly asymmetric systems ($q \lesssim 0.1$), whereas LGWA performs better in this regime, primarily due to its longer observation window and accumulating more SNR.
However, in Figures \ref{fig:5111} and \ref{fig:uniform}, we find ET outperforms LGWA and multi-band observations in recovering the mass ratio distribution, mainly because the ET detections are more fall in the low mass regime and the detection rate is less sensitive to variations in the mass ratio, as shown in Figure~\ref{fig:q_m}, while the LGWA and ET+LGWA has more detections in the high-mass regime, and the mass-ratio will affect the detection capability more in the high-mass regime.

Finally, regarding the redshift distributions (right panels), we convert the measured luminosity distance into redshift in the $\Lambda$CDM model with cosmological parameters fixed to the Planck 2018 result \cite{Planck:2018vyg}. 
As shown in the right panels of Figures~\ref{fig:2111}--\ref{fig:uniform}, in recovering the shape of the redshift distribution of IMBHs, three GW configurations show the similar performance. 

\section{Conclusion}\label{Conclusion}

IMBHs are widely regarded as the crucial link between stellar-mass and supermassive black holes. However, no definitive EM evidence has yet been found. GW observations provide a powerful means of directly detecting IMBH merger events. The GW signals from IMBH mergers primarily fall in the decihertz frequency band, where the lunar-based detector LGWA offers exceptionally high sensitivity, making it well suited to capture such signals. In contrast, the ground-based detector ET is more sensitive to the late merger phase of lower-mass systems, effectively compensating for LGWA's observational limitations in this region. Therefore, multi-band observations combining LGWA and ET can fully exploit their complementary strengths, enabling continuous tracking of the entire coalescence process of IMBH systems and significantly improving detection capabilities.

In this study, we first simulate the parameter distributions of IMBH binary sources based on three distinct population models. 
We then calculate the SNRs for each detector configuration and use the FIM to estimate parameter uncertainties, while assessing the effects of redshift, primary mass, inclination angle, and mass ratio on detectability.
Finally, we examine the detectable distributions for different detectors and explore the advantages of multi-band observations over single-detector observations in enhancing detection performance and recovering the intrinsic population distributions.

Our results demonstrate that multi-band observations combining LGWA and ET significantly enhance the detection capabilities for IMBH binary mergers across a broad parameter space. Within the joint network, ET's high sensitivity to low-mass and highly asymmetric mass-ratio systems compensates for LGWA's limitations in these regions, while LGWA excels at detecting high-mass systems. Together, they complement each other and effectively expand coverage across parameters such as primary mass, redshift, mass ratio, and inclination angle. 
Across three distinct population models, multi-band observations can better recover the intrinsic mass distribution and reduce the biases resulting from the limited detection capabilities of single detectors.
We conclude that multi-band GW synergetic observations play an essential role in detecting IMBHs and uncovering the population of their binary systems, offering a promising path toward a deeper understanding of IMBHs.

\section*{Acknowledgements}
This work was supported by the National Natural Science Foundation of China (Grants Nos. 12575049, 12533001, and
12473001), the National SKA Program of China (Grants Nos. 2022SKA0110200 and 2022SKA0110203), the China Manned Space Program (Grant No. CMS-CSST-2025-A02), and the 111 Project (Grant No. B16009).

\bibliographystyle{JHEP}
\bibliography{gwmultiband}

\end{document}